\documentstyle[11pt]{article}
\newlength{\bredde}
\def\slash#1{\settowidth{\bredde}{$#1$}\ifmmode\,\raisebox{.15ex}{/}
\hspace*{-\bredde} #1\else$\,\raisebox{.15ex}{/}\hspace*{-\bredde} #1$\fi}
\newcommand{\beq}{\begin{equation}}
\newcommand{\eeq}{\end{equation}}
\newcommand{\noi}{\vspace{12pt}\noindent}
\newcommand{\lG}{\raise.3ex\hbox{$\stackrel{\leftarrow}{G}$}}
\newcommand{\lU}{\raise.3ex\hbox{$\stackrel{\leftarrow}{U}$}}
\newcommand{\lP}{\raise.3ex\hbox{$\stackrel{\leftarrow}{{\cal P}}$}}
\newcommand{\leta}{\raise.3ex\hbox{$\stackrel{\leftarrow}{\eta}$}}
\newcommand{\lOmega}{\raise.3ex\hbox{$\stackrel{\leftarrow}{\Omega}$}}
\newcommand{\ldr}{\raise.3ex\hbox{$\stackrel{\leftarrow}{\delta^r}$}}

\newcommand{\Struve}{{\mbox{\bf L}}}
\def\beqn{\begin{eqnarray}}
\def\eeqn{\end{eqnarray}}

\def\gtwid{\raise.3ex\hbox{$>$\kern-.75em\lower1ex\hbox{$\sim$}}}
\def\ltwid{\raise.3ex\hbox{$<$\kern-.75em\lower1ex\hbox{$\sim$}}}

\def\la{\lambda}

\begin{document}
\topmargin -1.4cm
\oddsidemargin -0.8cm
\evensidemargin -0.8cm
\title{\Large{{\bf Massive Spectral Sum Rules for the Dirac Operator}}}

\vspace{1.5cm}

\author{~\\~\\
{\sc Poul H. Damgaard}\\
The Niels Bohr Institute\\ Blegdamsvej 17\\ DK-2100 Copenhagen\\
Denmark}
 
\maketitle
\vfill
\begin{abstract} Massive spectral sum rules are derived for Dirac operators
of $SU(N_c)$ gauge theories with $N_f$ flavors. The universal microscopic 
massive spectral densities of random matrix theory, where known, are all
consistent with these sum rules.
\end{abstract}
\vfill
\begin{flushleft}
NBI-HE-97-57 \\
hep-th/9711047
\end{flushleft}
\thispagestyle{empty}
\newpage

\setcounter{page}{1}

\noi
The derivation of exact spectral sum rules for eigenvalues of the Dirac
operator in QCD by Leutwyler and Smilga \cite{LS} has recently led to
a remarkable series of results. Most impressive are the developments
from random matrix theory. There is now quite compelling evidence that
the so-called microscopic spectral density of QCD (and hence, as a simple 
by-product also all spectral sum rules) can be computed
{\em exactly} from an appropriate large-$N$ matrix ensemble \cite{SV,V}.
The essential input is the existence of a non-vanishing spectral density
$\rho(\lambda)$ at the origin $\la \!=\! 0$, which, on account of the
Banks-Casher relation\footnote{
Here $N$ denotes the volume of space-time. We use this notation for
convenience, since it in matrix model language essentially corresponds to
the size of the random matrices.} $\rho(0) = 
N\Sigma/\pi~,~\Sigma \equiv \langle\bar{\psi}\psi\rangle$,
translates into the existence of a chiral condensate in QCD. {}From this
perspective the microscopic spectral densities embody information about
the symmetries of the Dirac operators alone, without any reference to
the detailed dynamics of the gauge field interactions. If correct, this
represents a major step towards the understanding and classification of
spontaneous chiral symmetry breaking in QCD and related theories. In
fact, the spectral sum rules of Leutwyler and Smilga are then seen to
be of one very particular universality class, out of many. The 
field-theoretic generalization of the sum rules to other patterns of
chiral symmetry breaking has been given in ref. \cite{SmV}. Furthermore,
in a very impressive series of papers \cite{SV,V} it has been shown
how to translate the (conjectured) catalog of universality classes of
random matrix models into gauge theory language (see, $e.g.$, ref. 
\cite{Vrev} for a recent review).

\noi
Evidence for the above scenario has been mounting after the initial
observation that massless spectral sum rules of QCD could be reproduced
by microscopic spectral densities computed in random matrix theory. 
First, it has been
shown that the microscopic spectral densities of random matrix theory
indeed do fall into distinct universality classes, independent of the
detailed matrix model interactions \cite{ADMN}. Second, there are now
direct numerical measurements of the microscopic spectral density for
quenched $SU(2)$ gauge theory, and the agreement with predictions from
random matrix theory is quite spectacular \cite{BBMSVW}. 

\noi
It was suggested by Shuryak and Verbaarschot \cite{SV} that the 
microscopic spectral density (and corresponding spectral sum rules),
which normally are considered for massless Dirac operators, could be
generalized to the massive case by an appropriate rescaling
of masses. Such an extension is in fact essential for future comparisons
with lattice gauge theory, where dynamical fermions in a finite volume
necessarily must be massive. Very recently, the associated double-microscopic
spectral densities (called so because both eigenvalues and masses are
rescaled as the volume $N$ is taken to infinity) have been derived from
random matrix theory, and proven to be universal as well \cite{DN}. A few
examples showed agreement with corresponding massive sum rules of QCD.

\noi
The purpose of this paper is to more systematically derive massive
spectral sum rules for massive Dirac operators, and to confront these sum 
rules with the recent results \cite{DN} from random matrix theory. 
Throughout we restrict ourselves to the region $1/\Lambda_{QCD}
\ll N^{1/4} \ll 1/m_{\pi}$. Here $\Lambda_{QCD}$ is a typical hadronic scale 
in QCD, and $m_{\pi}$ is the pion mass.

\noi
\underline{Gauge group $SU(N_c), N_c \geq 3, N_f$ fermions in the fundamental 
representation:}\\~\\
The coset is here $SU(N_f)_L\times SU(N_f)_R/SU(N_f)$, and the matrix
model ensemble is that of chiral unitary matrices \cite{Vrev}. Let us start
with the case $N_f\!=\! 1$, and the sector of zero topological charge: 
$\nu\!=\!0$. 
Massive spectral sum rules can now be derived in several ways.   
Consider the general expression for the
partition function $Z_v$ in the topological 
sector $\nu$ ($I_n(x)$ denotes the $n$th modified Bessel function) 
\cite{LS},
\beq
Z_{\nu} ~=~ I_{\nu}(\mu) ~,~~~~~ \mu \equiv N\Sigma m.
\eeq
{}From the expansion of the Bessel function, and a comparison with the 
corresponding formal expansion of the full QCD partition function it
follows that
\begin{eqnarray}
\frac{1}{N^2\Sigma^2} \left\langle \sum_n~\!' \frac{1}{\lambda_n^2 + m^2} 
\right\rangle_{\nu} &=& \sum_n \frac{1}{j_{\nu,n} + \mu^2} \cr
&=& \frac{I_{\nu+1}(\mu)}{2\mu I_{\nu}(\mu)} ~.\label{sum0}
\end{eqnarray}
where the $j_{\nu,n}$'s denote the zeros of the 
Bessel function $J_{\nu}(x)$, and where the zero mode has been omitted 
from the sum on the left hand side. 

\noi
The double-microscopic spectral densities 
$\rho_S^{(N_f)}(\la;\mu_1,\ldots,\mu_{N_{f}})$ are defined by
\beq
\rho_S^{(N_{f})}(\zeta;\mu_1,\ldots,\mu_{N_{f}}) 
~\equiv~ \lim_{N\to\infty} \frac{1}{\Sigma N}\rho\left(
\frac{\zeta}{\Sigma N}\right) ~, ~~~~~~~~ \mu_i = m_iN\Sigma ~~{\mbox{\rm   
fixed}}
\eeq
where $\rho(\la)$ is the ordinary, macroscopic, spectral density:
\beq
\rho(\lambda) ~=~ \langle \sum_n~\!' \delta(\lambda - \lambda_n)\rangle ~.
\eeq
With the help of this definition the sum rule (\ref{sum0}) can be written
\beq
\int_0^{\infty} \! d\zeta~ \frac{\rho_S^{(1)}(\zeta;\mu)}{\zeta^2 + \mu^2} ~=~
\frac{I_{\nu+1}(\mu)}{2\mu I_{\nu}(\mu)} ~.\label{sum1}
\eeq

\noi
We now compare this prediction with that of the chiral unitary random
matrix ensemble. 
In ref. \cite{DN} the derivation of the double-microscopic spectral density 
was restricted to the sector of zero topological charge, $\nu\!=\!0$. 
However, a simple analytical expression was also given for the case of
one massive and $N_f-1$ massless fermions, which conveniently can be
written\footnote{Analogous expressions valid for any $\nu$ for higher values 
of $N_f$ are contained in the general solution given in ref. \cite{DN} by 
considering the case of $\nu$ massless fermions and $N_f$ massive ones.} 
\begin{eqnarray}
\rho_S^{(N_f)}(\la;0,\ldots,0,\mu) &=& \frac{|\zeta|}{2}[J_{N_{f}}(\zeta)^2 -
J_{N_{f}-1}(\zeta)J_{N_{f}+1}(\zeta)] \cr &&+ \frac{|\zeta|\mu^2}{2N_{f}
(\zeta^2
+ \mu^2)}J_{N_{f}-1}(\zeta)\left[\frac{I_{N_{f}+1}(\mu)}{I_{N_{f}-1}(\mu)}
J_{N_{f}-1}(\zeta) + J_{N_{f}+1}(\zeta)\right] ~.
\end{eqnarray}
Comparing with the general matrix model expression \cite{V}, this 
immediately gives us also the double-microscopic spectral density for
$N_f\!=\! 1$ in the sector of arbitrary topological charge $\nu$:
\beq
\rho_S^{(1)}(\la;\mu) = \frac{|\zeta|}{2}[J_{\nu+1}(\zeta)^2 -
J_{\nu}(\zeta)J_{\nu+2}(\zeta)] + \frac{|\zeta|\mu^2 J_{\nu}(\zeta)}{2(\nu+1)
(\zeta^2\! +\! \mu^2)}
\left[\frac{I_{\nu+2}(\mu)}{I_{\nu}(\mu)}
J_{\nu}(\zeta) + J_{\nu+2}(\zeta)\right] ~.
\eeq
Substituting the above expression into the sum rule (\ref{sum1}), and 
performing the integral, one finds that this sum rule indeed is satisfied.

\noi
One notices that the spectral sum rule (\ref{sum1}) essentially is 
a rewriting of the (mass dependent) chiral condensate. The only slight
complication is that the sum rules conventionally are written in terms
on non-zero eigenvalues only. Because the partition function for one
massive flavor can be
written as ($\la_n > 0$ and, for convenience in all that follows, $\nu > 0$)
\beq
Z_{\nu} ~=~ \int [dA_{\mu}]_{\nu} ~m^{\nu}\prod_{n\neq 0}  (\la_n^2 + m^2)
\exp\{-S[A]\} ~,
\eeq
it follows that the general formula for the contribution of non-zero
modes to the mass-dependent chiral condensate $\Sigma(m)$ reads
\beq
\Sigma(m)_{\la\neq 0} ~=~ \Sigma\left[\frac{\partial}{\partial\mu}
\ln Z_{\nu}(\mu) - \frac{\nu}{\mu}\right] ~.\label{sigma1}
\eeq
In terms of the eigenvalues $\la_n$ it becomes
\beq
\frac{1}{N^2\Sigma^2}\left\langle \sum_n~\!' \frac{1}{\la_n^2 + m^2}
\right\rangle_{\nu} = \frac{1}{2\mu}\left[\frac{\partial}{\partial\mu}
\ln Z_{\nu}(\mu) - \frac{\nu}{\mu}\right] ~.\label{sumsigma}
\eeq
This formula, and generalizations for higher inverse moments, will give
us all the required massive spectral sum rules.\footnote{The restriction to
non-zero modes is not essential, and done here only in order to have
a well-defined massless limit.} For example, we
immediately recover the expression (\ref{sum1}) without having to
resort to the infinite summation over zeros of the Bessel function. It
is also worthwhile noting that the massless spectral sum rules of
Leutwyler and Smilga are recovered from the above expression in the massless
limit:
\beq
\frac{1}{N^2\Sigma^2} \left\langle \sum_n~\!' \frac{1}{\lambda_n^2} 
\right\rangle_{\nu} ~=~ \lim_{\mu\to 0}
\frac{I_{\nu+1}(\mu)}{2\mu I_{\nu}(\mu)} ~=~ \frac{1}{4(\nu+1)} ~.
\eeq
 
\noi
We now turn to the general case of $N_f \geq 2$. The partition function
$Z_{\nu}$ for equal masses $m = \mu/(N\Sigma)$ was found in the original 
work of ref. \cite{LS} as the determinant of an $N_f\!\times\! N_f$ matrix,
\beq
Z_{\nu} ~=~ \det M ~,~~~~~~~~~ M_{ij} ~=~ I_{\nu+j-i}(\mu) ~.
\eeq
More recently Jackson, \c{S}ener and Verbaarschot \cite{JSV} have
given the general formula for different masses:
\beq
Z_{\nu}(\mu_1,\ldots,\mu_{N_{f}}) = 
2^{N_{f}(N_{f}-1)/2}\left(\prod_{k=1}^{N_{f}}
(k-1)!\right) \frac{\det A}{\Delta(\mu^2)} ~,\label{znonequalm}
\eeq
where
\beq
A_{ij} ~=~ \mu_i^{j-1}I_{\nu}^{(j-1)}(\mu_i) 
\eeq
is given in terms of derivatives of the Bessel function, and
\beq
\Delta(\mu^2) ~=~ \prod_{i<j} (\mu_i^2 - \mu_j^2)
\eeq
is the Vandermonde determinant of masses.

\noi
For equal masses, the generalization of eq. (\ref{sumsigma}) to an
arbitrary number of flavors $N_f$ is
\beq
\frac{1}{N^2\Sigma^2}\left\langle \sum_n~\!' \frac{1}{\la_n^2 + m^2}
\right\rangle_{\nu} = \frac{1}{2N_f\mu}\left[\frac{\partial}{\partial\mu}
\ln Z_{\nu}(\mu) - N_f\frac{\nu}{\mu}\right] ~.\label{gensumsigma}
\eeq
To illustrate this, consider the case $N_f\!=\!2$, where this sum rule becomes
\beq
\left\langle \sum_n~\!' \frac{1}{\la_n^2 + m^2}
\right\rangle_{\nu} ~=~ \frac{N^2\Sigma^2}{4\mu}\left\{\frac{I_{\nu}(\mu)
I_{\nu+1}(\mu) - I_{\nu-1}(\mu)I_{\nu+2}(\mu)}{I_{\nu}(\mu)^2 - I_{\nu+1}(\mu)
I_{\nu-1}(\mu)}\right\} ~. \label{sumcomp2}
\eeq
As a simple by-product we recover the Leutwyler-Smilga sum rule for 
$N_f\!=\!2$ in the massless limit \cite{LS}:
\beq
\left\langle \sum_n~\!' \frac{1}{\la_n^2}
\right\rangle_{\nu} ~=~ \frac{N^2\Sigma^2}{4(\nu + N_f)} ~.\label{su3sr}
\eeq
The explicit expressions become increasingly involved with growing $N_f$.
For $N_f\!=\!3$, the sum rule analogous to (\ref{sumcomp2}) reads
\begin{eqnarray}
\left\langle \sum_n~\!' \frac{1}{\la_n^2 + m^2}
\right\rangle_{\nu} &~=~& \frac{N^2\Sigma^2}{6\mu Z_{\nu}}(I_{\nu}(\mu)^2
I_{\nu+1}(\mu) - I_{\nu+1}(\mu)^2I_{\nu-1}(\mu) + I_{\nu-2}(\mu)
I_{\nu+1}(\mu)I_{\nu+2}(\mu) \cr
&& - I_{\nu-1}(\mu)I_{\nu}(\mu)I_{\nu+2}(\mu)
+ I_{\nu-1}(\mu)^2I_{\nu+3}(\mu) - I_{\nu-2}(\mu)I_{\nu}(\mu)I_{\nu+3}(\mu))
~,\label{sumcomp3}
\end{eqnarray}
where
\beq
Z_{\nu} = I_{\nu}(\mu)^3 + I_{\nu+1}(\mu)^2I_{\nu-2}(\mu) + I_{\nu-1}(\mu)^2
I_{\nu+2}(\mu) - I_{\nu}(\mu)I_{\nu-2}(\mu)I_{\nu+2}(\mu)
-2I_{\nu-1}(\mu)I_{\nu}(\mu)I_{\nu+1}(\mu) ~.
\eeq
By explicitly taking the limit $\mu\to 0$ on the right hand side of eq.
(\ref{sumcomp3}), we also here confirm that it
correctly reduces to the massless sum rule (\ref{su3sr}) for $N_f\!=\!3$.

\noi
These massive sum rules can now be confronted with the predictions from
random matrix theory. Taking degenerate masses,
the double-microscopic spectral densities can be
written, for $\nu\!=\!0$: 
\beq
\rho_S^{(2)}(\zeta;\mu,\mu) = \frac{|\zeta|}{2}
\left[J_0(\zeta)^2 + J_1(\zeta)^2\right]
- \frac{2|\zeta|(\mu I_1(\mu)J_0(\zeta) + \zeta I_0(\mu)J_1(\zeta))^2}{(\zeta^2
+ \mu^2)^2[I_0(\mu)^2 - I_1(\mu)^2]} \label{rho2equal}
\eeq
for $N_f\!=\!2$. Using this density to evaluate the
left hand sides of eqs. (\ref{sumcomp2}), one
verifies that that this identity is satisfied. Even for equal masses, the 
expressions that must be integrated in order to verify the sum rules grow
rapidly with increasing $N_f$.

\noi
Many more different kinds of massive spectral sum rules can be derived
from the partition function for different quark masses (\ref{znonequalm}).
For example, for $N_f\!=\!2$  and $\nu\!=\!0$ we get, from eq. 
(\ref{znonequalm}),
\beq
\left\langle \sum_n~\!' \frac{1}{\la_n^2 + m_1^2}
\right\rangle_{\nu=0} ~=~ \frac{N^2\Sigma^2}{2\mu_1}\left\{
\frac{I_1(\mu_1)\mu_2I_1(\mu_2) - I_0(\mu_2)\mu_1I_0(\mu_1)}{I_0(\mu_1)
\mu_2I_1(\mu_2) - I_0(\mu_2)\mu_1I_1(\mu_1)} + \frac{2\mu_1}{\mu_2^2
- \mu_1^2}\right\} ~.
\eeq
We note that this correctly reduces to the sum rule (\ref{sumcomp2})
in the limit of degenerate fermion masses (cancellations are delicate, and
occur up to 2nd order). By numerical integration we have verified
to high accuracy that the microscopic spectral density for two fermion flavors
of different masses \cite{DN},
\begin{eqnarray}
\rho_S^{(2)}(\zeta;\mu_1,\mu_2) \! &=& \! \frac{|\zeta|}{2}\!\left(
J_0^2(\zeta) + J_1(\zeta)^2\right)  - \cr
&& \frac{|\zeta|(\mu_1^2 \!-\! \mu_2^2)}{(\zeta^2\!+\!\mu_1^2)
(\zeta^2\!+\!\mu_2^2)}
\frac{[\mu_1 I_1(\mu_1)J_0(\zeta) \!+\! \zeta I_0(\mu_1)J_1(\zeta)] 
[\mu_2 I_1(\mu_2)J_0(\zeta) \!+\! \zeta I_0(\mu_2)J_1(\zeta)]}
{\mu_1 I_1(\mu_1)I_0(\mu_2) - \mu_2 I_0(\mu_1)I_1(\mu_2)} ~.\label{r2}
\end{eqnarray}
is consistent with this sum rule. 

\noi
Massive spectral sum rules originating from higher derivatives of
$\ln Z_{\nu}$ with respect to (different) fermion masses can be derived
straightforwardly. Physically they are related to higher susceptibilities.
Using the general expression for the universal massive double-microscopic 
spectral {\em correlators} of ref. \cite{DN}, 
also these generalized sum rules are 
open to checks from random matrix theory. Higher massive sum rules for
different fermion masses satisfy non-trivial identities as one or more
of the masses are sent to infinity, due to decoupling. 
These identities have their direct
counterparts in the consistency conditions for $\rho_S$ and higher
correlators that were explained in ref. \cite{DN}. 

\noi
\underline{Gauge group $SU(N_c), N_c \geq 2, N_f$ fermions in the adjoint 
representation:}\\~\\
The coset is $SU(N_f)/SO(N_f)$, and the relevant random matrix models 
belong belong
to the chiral symplectic ensemble \cite{Vrev}. Massless spectral sum rules 
have been been derived by Leutwyler and Smilga for $N_f\!=\!1$ and 2 
\cite{LS}, and
later generalized to arbitrary $N_f$ by Smilga and Verbaarschot \cite{SmV}:
\beq
\left\langle \sum_n~\!' \frac{1}{\la_n^2}
\right\rangle_{\nu} ~=~ \frac{N^2\Sigma^2}{4(\bar{\nu} + (N_f + 1)/2)} ~.
\label{masslesssunadjoint}
\eeq
Here $\bar{\nu} \!\equiv\! N_c\nu$ is the number of zero modes, and on the left
hand side the sum runs over the doubly-degenerate, positive eigenvalues,
counted once. The partition function has been given explicitly for
$N_f\!=\!1$ \cite{LS}, where  $Z_{\nu} = I_{\bar{\nu}}(\mu)$. The massive
spectral sum rules therefore coincide with those of $SU(N_c\!\geq\!3)$
for $N_f\!=\!1$ and $\nu$ replaced by $\bar{\nu}$. In particular,
eq. (\ref{sum0}), with
$\nu$ replaced by $\bar{\nu}$, for the simplest of these massive sum rules. 
Also the partition function for $N_f\!=\!2$ and degenerate masses
is known \cite{LS}:
\beq
Z_{\nu} ~=~ \sum_{n=0}^{\infty}\frac{\mu^{2n+2\bar{\nu}}}{n!(n+2\bar{\nu})!
(2n+2\bar{\nu}+1)} ~.\label{zsunnadjoint}
\eeq
After some manipulations, we have simplified this expression to
\beq
Z_{0} = I_0(2\mu) + \frac{\pi}{2}\left[I_0(2\mu)
\Struve_1(2\mu) - I_1(2\mu)\Struve_0(2\mu)\right] 
\eeq
for $\bar{\nu}\!=\!0$, and
\beq
Z_{\nu} = (-1)^{\bar{\nu}}\left\{I_0(2\mu) + \frac{\pi}{2}\left[I_0(2\mu)
\Struve_1(2\mu) - I_1(2\mu)\Struve_0(2\mu)\right] -
\frac{1}{\mu}\sum_{k=0}^{\bar{\nu}-1}
(-1)^kI_{2k+1}(2\mu)\right\}\label{sunadjointz}
\eeq
for $\bar{\nu}\geq 1$. Here $\Struve_n(x)$ is the $n$th 
modified Struve function.
We note that the right hand side of eq. (\ref{sunadjointz}), despite
appearances, is positive also for odd $\bar{\nu}$.

\noi
Massive spectral sum rules can now be derived by straigthforward 
differentiation, but for large values of $\bar{\nu}$ they yield rather
unwieldy expressions. We restrict ourselves here to displaying only the
simplest sum rule for $\bar{\nu}\!=\!0$, which in fact simplifies
considerably:
\beq
\left\langle \sum_n~\!' \frac{1}{\la_n^2 + m^2}\right\rangle_{\nu=0}
~=~ \frac{N^2\Sigma^2\pi}{8\mu^2 Z_0(\mu)}\left[I_1(2\mu)\Struve_0(2\mu)
- I_0(2\mu)\Struve_1(2\mu)\right] ~.\label{masssunadjoint}
\eeq
Using the expansions $\Struve_0(x) = 2x/\pi + \ldots$ and $\Struve_1(x)
= 2x^2/(3\pi) + \ldots$, one verifies that this massive spectral sum rule
reduces to the massless spectral sum rule  (\ref{masslesssunadjoint})
in the limit $\mu\! \to\! 0$. There are presently no predictions from
random matrix theory with which to compare the massive spectral sum rule
(\ref{masssunadjoint}).

\noi
Staggered fermions of $SU(2)$ lattice gauge theory have the symmetries 
of the chiral symplectic matrix ensemble \cite{Vrev}. The massive sum rules
for $N_f\!=\!1$ and $N_f\!=\!2$ shown above may therefore soon be tested by 
Monte Carlo simulations. 

\noi
\underline{Gauge group $SU(2), N_f$ fermions in the fundamental 
representation:}\\~\\
The coset is here $SU(2N_f)/Sp(2N_f)$, and the matrix
models are of the chiral orthogonal ensemble \cite{Vrev}. The massless
spectral sum rules have been derived by Smilga and Verbaarschot \cite{SmV},
and found, for the simplest case, to be of the form
\beq
\left\langle \sum_n~\!' \frac{1}{\la_n^2}
\right\rangle_{\nu} ~=~ \frac{N^2\Sigma^2}{4(\nu + 2N_f - 1)} ~.
\label{masslesssu2}
\eeq
The partition function $Z_{\nu}$ in the sector of topological charge $\nu$
can, for $N_f$ flavors of equal masses $m\!=\! \mu/(N\Sigma)$, be written
in terms of a Pfaffian \cite{SmV}:
\beq
Z_{\nu} ~=~ \frac{1}{(2N_f-1)!!} Pf(A) 
~,~~~~~ A_{ij} ~=~ (j-i)I_{i+j+1}(\mu) ~.\label{su2fund}
\eeq

\noi
For $N_f\!=\!1$ this formula gives $Z_{\nu} = I_{\nu}(\mu)$, and the
massive spectral sum rules therefore coincide with those of $SU(N_f\geq 3),
N_f = 1$, and in particular eq. (\ref{sum0}) for the simplest sum rule.
For $N_f\!=\!2$ we get, from eqs. (\ref{gensumsigma}) and (\ref{su2fund}),
\beq
\left\langle \sum_n~\!' \frac{1}{\la_n^2 + m^2}
\right\rangle_{\nu} ~=~ \frac{N^2\Sigma^2}{2\mu}\left\{\frac{
2I_{\nu}(\mu)I_{\nu+1}(\mu) - 3I_{\nu-1}(\mu)I_{\nu+2}(\mu)
+ I_{\nu-2}(\mu)I_{\nu+3}(\mu)}
{3I_{\nu}(\mu)^2 
- 4I_{\nu-1}(\mu)I_{\nu+1}(\mu) + I_{\nu-2}(\mu)I_{\nu+2}(\mu)}\right\} ~. 
\label{sumcomp2su2}
\eeq
As a direct check of this expression, we also here find that it reduces 
to the known result (\ref{masslesssu2}) in the massless limit $\mu \to 0$.
Expressions for larger values of $N_f$ can be worked out analogously.
There are presently no predictions from random matrix theory with which to
compare these massive spectral sum rules.
\vspace{0.1cm}

\noi
We end this paper with a few comments on the physical significance of
these massive sum rules. The whole idea is to rescale both eigenvalues
$\la$ and fermion masses $m_i$ at the same rate as the volume is taken
to infinity, so that $\zeta\!\equiv\!\la N\Sigma$ and $\mu_i\!\equiv\!
m_i N\Sigma$ are kept fixed. In addition, we are restricted to volumes
$N$ that satisfy  $1/\Lambda_{QCD} \ll N^{1/4} \ll 1/m_{\pi}$. To obtain a 
chiral condensate of the theory with massless fermions in the infinite-volume 
limit, it is well-known that one should {\em first} take the volume $N$ to 
infinity, and then take $m_i$ to zero. This is not the limit considered here, 
where the $m_i\!\to\! 0$ at precisely just the proper rate with $N\!\to\!
\infty$ to leave a well-defined double-microscopic massive
spectral density. Highly non-trivial
information about the finite-volume correlations of eigenvalues is
contained in this limiting function. From the point of view of lattice
gauge theory simulations, the limit we have discussed here is completely
realizable. The required tuning of quark masses
as different lattice volumes are considered is an entirely natural
and feasible operation in Monte Carlo simulations. There are
therefore good reasons to believe that the massive spectral sum rules discussed
here may be tested in lattice gauge theory. It is already a highly
non-trivial fact that the double-microscopic massive spectral densities
of random matrix theory, in those cases where known, are consistent
with these sum rules.


\end{document}